\documentclass[reprint,superscriptaddress,amssymb,amsmath,aps,showpacs,10pt,longbibliography,prb]{revtex4-2}
\usepackage{graphicx}

\graphicspath{{figures/}}
\usepackage{color}
\usepackage[colorlinks,bookmarks=false,citecolor=darkblue,linkcolor=red,urlcolor=blue]{hyperref}

\definecolor{darkred}{rgb}{0.7,0.0,0.0}
\definecolor{darkblue}{rgb}{0,0.02,0.45}

\definecolor{darkgreen}{rgb}{0.02,0.45,0.0}

\begin{document}


\title{Revisiting the symmetry and optical phonons of altermagnetic $\alpha$-MnTe}

\author{Ece Uykur}
\email{e.uykur@hzdr.de}
\affiliation{Helmholtz-Zentrum Dresden-Rossendorf, Inst Ion Beam Phys \& Mat Res, D-01328 Dresden, Germany}

\author{Marcos V. Gon\c{c}alves-Faria}
\affiliation{Helmholtz-Zentrum Dresden-Rossendorf, Inst Ion Beam Phys \& Mat Res, D-01328 Dresden, Germany}

\author{Sahana R\"o{\ss}ler}
\author{Victoria A. Ginga}
\affiliation{Felix Bloch Institute for Solid-State Physics, University of Leipzig, 04103 Leipzig, Germany}

\author{Marcus Schmidt}
\affiliation{Max Planck Institute for Chemical Physics of Solids, 01067 Dresden, Germany}

\author{Stephan Winnerl}
\affiliation{Helmholtz-Zentrum Dresden-Rossendorf, Inst Ion Beam Phys \& Mat Res, D-01328 Dresden, Germany}

\author{Manfred Helm}
\affiliation{Helmholtz-Zentrum Dresden-Rossendorf, Inst Ion Beam Phys \& Mat Res, D-01328 Dresden, Germany}

\author{Alexander A. Tsirlin}
\email{altsirlin@gmail.com}
\affiliation{Felix Bloch Institute for Solid-State Physics, University of Leipzig, 04103 Leipzig, Germany}


\begin{abstract}
Using infrared (IR) and Raman spectroscopies combined with high-resolution x-ray diffraction, we address several controversial aspects of altermagnetic $\alpha$-MnTe. We show that mechanical stress applied to crystals of this material causes a drastic broadening of Bragg peaks that conceals signatures of additional phases present in the sample. Indeed, spatially resolved Raman spectroscopy reveals that the modes around 175\,cm$^{-1}$ often reported in $\alpha$-MnTe are not reproducible across different positions and samples and originate from the secondary phase of MnTe$_2$. By combining spectroscopic probes with \textit{ab initio} calculations, we establish the IR-active optical phonon of $\alpha$-MnTe around 155\,cm$^{-1}$ ($E_{1u}$) and the Raman-active optical phonon around 100\,cm$^{-1}$ ($E_{2g}$) at room temperature. Two intense Raman modes around 120 and 140\,cm$^{-1}$ are shown to be intrinsic, even though they can not be assigned to $\Gamma$-point optical phonons. These modes couple to magnetic order in $\alpha$-MnTe and also to the transient reflectivity resulting in coherent oscillations. Both 6-fold rotation symmetry and inversion symmetry are preserved in $\alpha$-MnTe within our experimental resolution. 
\end{abstract}

\maketitle


\section{Introduction}
Altermagnets have drawn significant attention in recent years due to their unique properties \cite{smejkal2022a, smejkal2022b}. Microscopically, they exhibit an antiparallel spin configuration analogous to that of antiferromagnets, resulting in a vanishing net magnetization. Simultaneously, they are characterized by the presence of spin-split electronic bands -- a canonical feature of ferromagnets -- permitting the generation and manipulation of spin-polarized electric currents. Adapting the best of both worlds, ferromagnets and antiferromagnets, altermagnets constitute an emergent and technologically promising subclass of magnetic materials \cite{bai2024, fender2025}.

Among the proposed candidates, $\alpha$-MnTe is a confirmed $g$-wave altermagnet that has been studied in bulk as well as thin-film forms~\cite{kriegner2016,kriegner2017,betancourt2023,betancourt2024,kluczyk2024,hariki2024,hubert2025}. It crystallizes in the NiAs-type structure with the space group $P6_3/mmc$. It is a robust collinear antiferromagnet below $T_N\sim307$~K~\cite{uchida1956,kunitomi1964}. Both the thin films and single crystals of this compound exhibit spin-split electronic bands that were confirmed in recent angle-resolved photoemission spectroscopy experiments~\cite{lee2024,krempasky2024,hajlaoui2024,osumi2024}. Concurrently, the splitting of chiral magnon bands was demonstrated by inelastic neutron scattering~\cite{liu2024}. 

In view of the confirmed room-temperature altermagnetic behavior of $\alpha$-MnTe, it is desirable to develop an experimental control of the altermagnetic order parameter and the associated anomalous Hall response of the material~\cite{mazin2023,mazin2024}. Whereas mechanical stress can be useful in this regard~\cite{liu2025,smolenski2025}, light offers an interesting alternative. In this context, understanding phonon and magnon excitations and their coupling to light is of importance. As a simple binary compound, $\alpha$-MnTe has been a subject of several spectroscopic studies, including Raman and infrared (IR) spectroscopy experiments, but they have produced conflicting results. 

Raman spectra of $\alpha$-MnTe display several modes. The one around 175\,cm$^{-1}$ is often indicated as the Raman-active $E_{2g}$ phonon~\cite{mobasser1985, szuszkiewicz1997, zhang2020}, whereas two further modes around 120 and 140\,cm$^{-1}$ are generally attributed to an impurity phase of elemental tellurium~\cite{mueller1991,szuszkiewicz2014,shao2026}. Interestingly, all three modes produce coherent phonon oscillations. The 175\,cm$^{-1}$ one appears in the ultrafast magneto-optical Kerr effect (MOKE) measurements performed with the 800\,nm excitation energy~\cite{bossini2021}, which exceeds the band gap of $\alpha$-MnTe. On the other hand, experiments with an in-gap excitation energy (1560\,nm wavelength) revealed similar oscillations for the 120 and 140\,cm$^{-1}$ modes~\cite{gray2024}. 
It was also suggested that the 175\,cm$^{-1}$ excitation is in fact a silent mode that appears in the Raman spectrum because of a subtle structural distortion lifting inversion symmetry of the crystal structure~\cite{wu2025}. Alternatively, the possible plasmonic origin of this mode was discussed~\cite{thapa2026}. 

IR studies of $\alpha$-MnTe consistently show the band gap of about 1.4\,eV~\cite{allen1977,ferrerRoca2000,bossini2020} and a phonon excitation in the $130-154$\,cm$^{-1}$ range~\cite{povstyanyi1972, onari1974, allen1977, gao2026}, although the nature of this phonon has not been established. Additionally, an optical magnon mode was observed at 28\,cm$^{-1}$ ($\sim$3.5~meV) in a recent magneto-transmission study in the THz region~\cite{dzian2025}, consistent with inelastic neutron scattering observations. The second magnon mode of an easy-plane antiferromagnet resides at zero energy and becomes visible only when magnetic field is applied~\cite{povarov2025}.

Seeking to resolve these ambiguities of the optical phonon modes, we performed a systematic study of \mbox{$\alpha$-MnTe} using both IR and Raman spectroscopies combined with high-resolution x-ray diffraction (XRD) as an independent probe of the structural symmetry. 
By tracking temperature and spatial dependence of individual excitations, we show that the 175\,cm$^{-1}$ Raman mode arises from the secondary phase of MnTe$_2$. On the other hand, the 120 and 140\,cm$^{-1}$ Raman modes, previously thought to be extrinsic, in fact belong to $\alpha$-MnTe and likely arise from zone-boundary phonons. Our data show no signatures of a symmetry lowering in $\alpha$-MnTe either above or below $T_N$. 



\section{Methods}

Single crystals of $\alpha$-MnTe were grown by chemical transport, as described in Ref.~\cite{roessler2025}. Crystals from two different batches were used in this study. Additionally, we performed XRD measurements on a polycrystalline sample of $\alpha$-MnTe that was prepared by annealing stoichiometric amounts of Mn and Te and  served as the source for the crystal growth~\cite{roessler2025}. Resistivity measurements were performed on the single crystals from both batches in zero magnetic field using Physical Property Measurement System (Quantum Design) in the temperature range of $380-2$\,K upon cooling. The electrical contacts were made in linear geometry using Au-wires and silver epoxy. The samples from the two batches (S1 and S2) show comparable absolute values of the resistivity but a different temperature dependence, with a sharper initial drop of the resistivity in S2 (see also Sec.~\ref{sec:raman}). 

High-resolution XRD data were collected at the ID22 beamline of ESRF, Grenoble, using the x-ray wavelength of 0.4\,\r A and multi-analyzer detector setup~\cite{fitch2023}. The samples were placed into thin borosilicate-glass capillaries and spun during the measurement. An agate mortar was used for grinding. Sample temperature was controlled by the liquid-nitrogen cryostream. \texttt{Jana2006} program was used for the structure refinement~\cite{jana2006}.

Frequencies of $\Gamma$-point phonons were calculated within density functional theory (DFT) using the \texttt{VASP} code~\cite{vasp1,vasp2} and finite-displacement method. Several exchange-correlation functionals were tested, including PBE~\cite{pbe96}, PBE for solids (PBEsol)~\cite{pbesol}, and r$^2$SCAN~\cite{scan}, which is referred to as SCAN in the following. All calculations were performed for the ground-state magnetic configuration of $\alpha$-MnTe, with ferromagnetic layers antiferromagnetically ordered along the $c$-direction~\cite{kunitomi1964}. Additionally, we performed phonon calculations for MnTe$_2$ within its experimental non-collinear magnetic structure~\cite{burlet1997}. Experimental lattice parameters at room temperature were used in all calculations. The $12\times 12\times 6$ and $8\times 8\times 8$ $k$-mesh were applied for $\alpha$-MnTe and MnTe$_2$, respectively.

Raman spectroscopy measurements were performed on single crystals in quasi-backscattering geometry with a Horiba spectrometer with light polarization in the $ab$-plane. Temperature-dependent spectra were collected between 10 and 350~K with the $\lambda=532$~nm excitation energy. Samples were placed into a Janis cryostat, and measurements were performed in the micro-Raman configuration with a 50x objective. The spot size was $\sim$5~$\mu$m. At room temperature, we also performed polarization-dependent measurements in the co-polarized ($xx$) and cross-polarized ($xy$) configurations to further clarify the modes. 

An IR study was performed on a Bruker Vertex 80v spectrometer equipped with a reflection unit in the reflection geometry. An as-grown sample with a flat, shiny surface was placed into a Cryovac cryostat, and the spot size of $\sim$ 3 mm was used for the measurements. The spectra between 70 and 500\,cm$^{-1}$ in the temperature range of $10-330$~K were measured with a gold mirror serving as reference. Optical magnons in the THz region were investigated using a home-built THz-TDS (time-domain spectroscopy) system. The $0.2-3$ THz range was measured on a $\sim$100~$\mu$m thick sample in the transmission configuration. These complementary low-energy results are in a good agreement with the IR spectroscopy measurements at higher energies. Due to the partially transparent nature of the sample, the low-energy range can be probed in the transmission geometry, whereas the phonon absorption at higher energies can be traced with the reflection geometry. Both reflection and transmission are modeled for the partially transparent sample in the slab geometry.

Optical pump-probe experiments were performed in the reflection geometry at room temperature. We used 60~fs-long laser pulses, centered at 800~nm, generated by a Ti:sapphire laser amplifier with a 250 kHz repetition rate, for both the pump and probe beams.  The probe spot on the sample surface was around 25 $\mu$m (FWHM), while the pump spot was 35 $\mu$m (FWHM). Transient reflectivity was measured up to 150~ps delay time with 1~ps time resolution. For delays up to around 9~ps, we increased the time resolution to 33~fs to resolve the phonon oscillations.


\section{Results}

\subsection{Crystal structure}
In-plane direction of the magnetic moments in $\alpha$-MnTe breaks the 6-fold rotation symmetry and potentially facilitates a symmetry lowering of the crystal structure. Therefore, we used synchrotron powder XRD that offers the ultimate angular resolution to detect possible deviations from the hexagonal symmetry. The multi-analyzer detector setup can be used to reveal even minuscule changes in the lattice metric, but it also exposes any intrinsic broadening of the Bragg peaks caused by strain. This happens to be the case in $\alpha$-MnTe where measurement on a crushed single crystal revealed relatively broad peaks with the full-width-at-half-maximum (FWHM) of $0.045^{\circ}$ for the 101 reflection at $2\theta=7.23^{\circ}$, compared to the instrumental resolution of about $0.003^{\circ}$ at this scattering angle. 

\begin{figure*}
\includegraphics{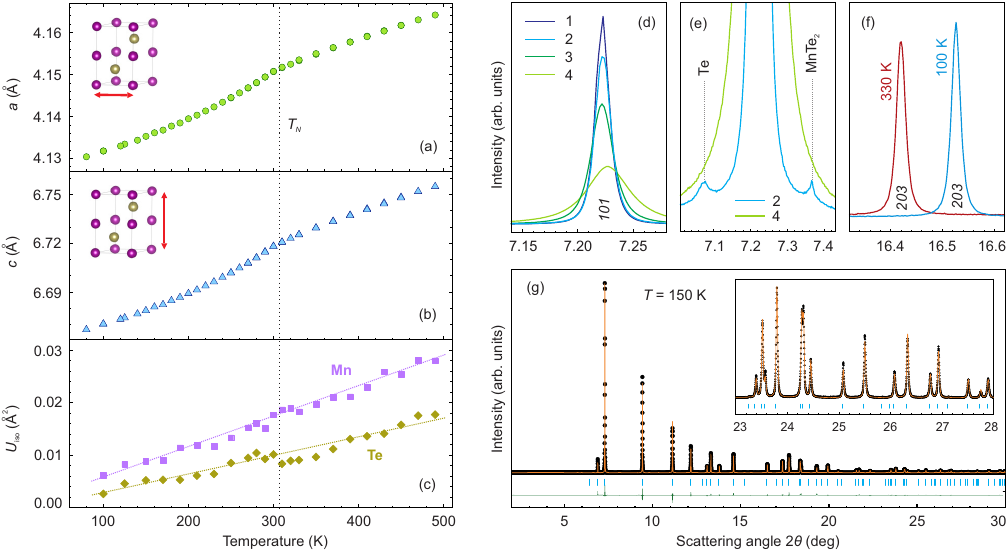}
\caption{\label{fig:xrd}
High-resolution XRD results for $\alpha$-MnTe. (a,b) Temperature dependence of the lattice parameters shows a magnetoelastic effect at $T_N$. (c) Temperature dependence of the atomic displacement parameters $U_{\rm iso}$, the lines are guides for the eye. (d) 101 Bragg peak measured on samples with different level of grinding, increasing from 1 to 4: non-ground polycristalline sample (1), sample gently ground for 30\,sec (2), sample gently ground for 3\,min (3), crushed and thoroughly ground single crystal (4). (e) Magnified view of the base of the same 101 Bragg peak reveals tiny impurity peaks of MnTe$_2$ and Te that are observable in a gently ground sample only. (f) 203 Bragg peak in sample 2 measured at 100\,K and 330\,K shows persistent hexagonal symmetry below $T_N$. (g) Rietveld refinement for the XRD data collected at 150\,K (sample 2). Tick marks show the Bragg peak positions for $\alpha$-MnTe.
}
\end{figure*}

Post-annealing the powder prepared by crushing the single crystal did not improve the peak width. A more fruitful approach turned out to be the use of the fresh $\alpha$-MnTe polycrystalline sample prepared as the starting material for the crystal growth~\cite{roessler2025}. Such a sample contains large crystalline grains and produces sharp XRD peaks (FWHM of $0.010^{\circ}$ for 101) when loaded into a capillary without any grinding. Attempts to grind this sample led to a rapid broadening of the peaks (Fig.~\ref{fig:xrd}d), demonstrating that $\alpha$-MnTe is highly sensitive to a mechanical stress. The large peak width observed for the the crushed single crystal should then be ascribed to the effect of grinding and not to the quality of the crystal itself. 

One downside of using the pristine (non-ground) polycrystalline sample in a powder XRD experiment is the insufficient powder averaging of the reflection intensities due to the presence of large crystallites. The best results in terms of the low FWHM and good powder averaging could be achieved by performing a very short (30\,sec), gentle grinding and using the fast spinning rate of 1000\,rpm during the data collection. The resulting XRD patterns allowed the complete structure refinement with the refinement residuals of $R_I=0.045-0.050$ (Fig.~\ref{fig:xrd}g). For comparison, we achieved $R_I=0.022$ in the thoroughly ground sample with the broadened peaks and good powder averaging. 

Before turning to the results of the structure refinement, we note that the reduced peak width in the polycrystalline sample facilitates the observation of tiny reflections due to secondary phases that can be otherwise concealed by the tails of the broad peaks if the sample is thoroughly ground (Fig.~\ref{fig:xrd}e). We identify 1.6(1)\,wt.\% of MnTe$_2$ and 0.5(1)\,wt.\% of Te in the polycrystalline sample used in this study. 
Traces of MnTe$_2$ may also occur in single crystals, as we show in the following.

\begin{table*}
\caption{\label{tab:mnte2}
Calculated phonon frequencies for MnTe$_2$ (in cm$^{-1}$) and the relevant experimental frequencies given as reference.
}
\begin{ruledtabular}
\begin{tabular}{lc@{\hspace{0.4cm}}cc@{\hspace{0.4cm}}c@{\hspace{0.4cm}}cc@{\hspace{0.4cm}}ccc@{\hspace{0.4cm}}ccccc}
                              & $A_g$ & $A_u$ &  & $E_g$ & $E_u$ & & $T_g$ &   &    & $T_u$ & & & \smallskip\\
PBE, $U_{\rm eff}=0$\,eV      &  130  & 129 & 62 &  91  & 128 & 57 & 144 & 93  & 83 & 138 & 133 & 124 & 74 & 55 \\
PBE, $U_{\rm eff}=2$\,eV      &  155  & 135 & 62 & 101  & 147 & 52 & 162 & 103 & 91 & 152 & 148 & 137 & 70 & 52 \\
PBE, $U_{\rm eff}=4$\,eV      &  164  & 140 & 61 & 106  & 160 & 50 & 169 & 107 & 94 & 163 & 157 & 146 & 67 & 51\smallskip\\
PBEsol, $U_{\rm eff}=0$\,eV   &  122  & 125 & 62 &  88  & 120 & 58 & 140 & 89  & 79 & 133 & 126 & 116 & 74 & 56 \\
PBEsol, $U_{\rm eff}=2$\,eV   &  154  & 133 & 61 &  99  & 141 & 52 & 159 & 100 & 87 & 147 & 143 & 132 & 70 & 53 \\
PBEsol, $U_{\rm eff}=4$\,eV   &  164  & 137 & 60 & 104  & 154 & 50 & 167 & 104 & 91 & 158 & 153 & 141 & 68 & 52\smallskip\\
SCAN, $U_{\rm eff}=0$\,eV     &  165  & 126 & 66 & 103  & 126 & 34 & 172 & 106 & 94 & 134 & 129 & 124 & 68 & 46 \\
SCAN, $U_{\rm eff}=2$\,eV     &  175  & 130 & 66 & 108  & 148 & 49 & 179 & 110 & 97 & 153 & 148 & 136 & 73 & 51 \\
SCAN, $U_{\rm eff}=4$\,eV     &  180  & 129 & 64 & 112  & 161 & 53 & 184 & 113 & 100 & 165 & 158 & 142 & 73 & 44\smallskip\\
Experiment~\cite{onari1974,mueller1991} & 177 & & & 128 & &        & 181 & 106 & 96 & 164 & 156 & 140 & 149 & \\
\end{tabular}
\end{ruledtabular}
\end{table*}

Our high-resolution XRD data do not reveal any peak splitting or broadening across $T_N$ and down to the lowest measured temperature of 80\,K. A large magnetoelastic effect is observed at $T_N$ (Fig.~\ref{fig:xrd}), in agreement with the previous studies~\cite{gronvold1972,baral2023}. Notwithstanding this fact, none of the reflections split, the lattice remains hexagonal, and the structure is consistent with the $P6_3/mmc$ symmetry at all temperatures. Another possible scenario is a symmetry lowering within the hexagonal crystal system, such as the loss of the inversion center discussed in Ref.~\cite{wu2025}. It would cause opposite shifts of the Mn atoms along the $c$-direction and appear as static disorder in the Mn position when the centrosymmetric space group $P6_3/mmc$ is used. Our data indeed show a relatively high atomic displacement parameter of Mn, $U_{\rm iso}\simeq 0.015$\,\r A$^2$ at room temperature. Importantly, this value systematically decreases upon cooling and essentially vanishes toward 0\,K (Fig.~\ref{fig:xrd}c). Therefore, the Mn displacements are dynamic in nature, and no indications for the lower symmetry, such as $P\bar 6m2$ proposed in Ref.~\cite{wu2025}, can be inferred from our data. 


\subsection{Calculation of phonon frequencies}

The interpretation of phonon modes in the IR and Raman spectra is typically assisted by DFT calculations. Therefore, it is important to establish a reliable computational procedure for obtaining accurate frequencies of optical phonons. To this end, we first consider MnTe$_2$, a compound with the same constituent elements, and compare its calculated phonon frequencies with the available experiments~\cite{onari1974,mueller1991}. With 12 atoms in the unit cell, MnTe$_2$ features 33 optical modes as follows:%
\begin{align*}
{\rm IR}\!:    & \quad 5\,T_u \smallskip\\
{\rm Raman}\!: & \quad A_g + E_g + 3\,T_g
\end{align*}%
whereas $2A_u$ and $2E_u$ modes remain silent. 

Different DFT functionals, ranging from standard GGA (generalized gradient approximation)~\cite{pbe96,pbesol} to the improved meta-GGA ones~\cite{scan}, can be used when calculating the phonons. Additionally, correlation effects in the Mn $3d$ shell can be taken into account on the mean-field DFT+$U$ level by introducing the on-site Coulomb repulsion parameter $U_{\rm eff}$~\footnote{Note that we use $U_{\rm eff}=U-J$ where $U$ and $J$ are the on-site Coulomb repulsion and Hund's coupling of DFT+$U$, respectively. We have verified that setting, for example, $U=5$\,eV and $J=1$\,eV vs. $U=4$\,eV and $J=0$\,eV produces essentially the same frequencies for all phonons.}. All these settings affect the phonon frequencies and cause a large spread of the calculated values, as shown in Table~\ref{tab:mnte2}. Several trends can be distinguished. First, PBEsol leads to a systematic softening of the phonons compared to PBE. Second, correlation effects generally cause phonon hardening. Third, meta-GGA hardens the even modes while leaving the odd modes mostly unchanged compared to the GGA results. 

\begin{table}
\caption{\label{tab:mnte}
Calculated phonon frequencies for $\alpha$-MnTe (in cm$^{-1}$).
}
\begin{ruledtabular}
\begin{tabular}{lcccccc}
                              & $B_{2u}$ & $A_{2u}$ & $B_{1g}$ & $E_{1u}$ & $E_{2g}$ & $E_{2u}$ \smallskip\\
PBE, $U_{\rm eff}=0$\,eV      &   161    &   121    &   116    &   104    &    86    &    71    \\
PBE, $U_{\rm eff}=2$\,eV      &   185    &   132    &   123    &   129    &    93    &    92    \\
PBE, $U_{\rm eff}=4$\,eV      &   201    &   140    &   127    &   142    &    98    &   104\smallskip\\

PBEsol, $U_{\rm eff}=0$\,eV   &   154    &   114    &   110    &    96    &    82    &    63    \\
PBEsol, $U_{\rm eff}=2$\,eV   &   180    &   126    &   118    &   123    &    90    &    87\\
PBEsol, $U_{\rm eff}=4$\,eV   &   185    &   129    &   120    &   127    &    91    &    90\smallskip\\

SCAN, $U_{\rm eff}=0$\,eV     &   163    &   110    &   127    &   119    &    93    &    70    \\
SCAN, $U_{\rm eff}=2$\,eV     &   187    &   125    &   131    &   143    &    98    &   107    \\
SCAN, $U_{\rm eff}=4$\,eV     &   205    &   137    &   134    &   157    &   102    &   126    \\ 
\end{tabular}
\end{ruledtabular}
\end{table}

The best agreement with the experimental phonon frequencies is obtained using SCAN with $U_{\rm eff}=4$\,eV. Most of the experimental frequencies of MnTe$_2$ are reproduced, with deviations below 5\,cm$^{-1}$ and only two discrepancies as follows. First, the experimental IR mode observed at 149\,cm$^{-1}$ is likely an overtone of the lower-energy $T_u$ phonon, because only three IR modes are expected in the $140-170$\,cm$^{-1}$ range. Second, the origin of the 128\,cm$^{-1}$ mode assigned to an $E_g$ phonon in Ref.~\cite{mueller1991} is unclear, but another, unexplained mode of the same symmetry appears in their data at $105-110$\,cm$^{-1}$ and would be in a much better agreement with the DFT results shown here.

The calculations for $\alpha$-MnTe show similar trends. With 4 atoms in the unit cell, one expects 9 optical modes, with $A_{2u}$ and $E_{1u}$ being IR-active polarized along the $z$- and $x,y$-directions, respectively, $E_{2g}$ being Raman-active (observed in both $xx-yy$ and $xy$ polarizations), whereas the remaining 4 modes ($B_{1g}$, $B_{2u}$, and $E_{2u}$) are silent. Table~\ref{tab:mnte} illustrates a large spread of the calculated values. Using MnTe$_2$ as reference, we posit that the most accurate results should be those obtained using SCAN with $U_{\rm eff}=4$\,eV. This is in agreement with our experimental results shown below.


\subsection{Raman response and the coherent phonons}
\label{sec:raman}

\begin{figure}
\includegraphics{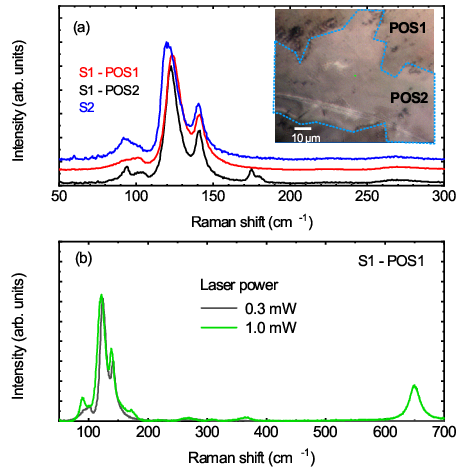}
\caption{\label{fig:R} (a) Representative Raman spectra for two $\alpha$-MnTe crystals. The inset shows the crystal S1, where two different regions, POS1 and POS2, can be distinguished. S1-POS2 is where we observe the 175 and 180\,cm$^{-1}$ modes in addition to the regular modes of $\alpha$-MnTe. (b) Background-subtracted spectra measured on S1-POS1 with two different laser powers. The 175\,cm$^{-1}$ mode along with several higher-energy modes appear when higher laser power is used.}
\end{figure}

\begin{figure}
\includegraphics[width=8.6cm]{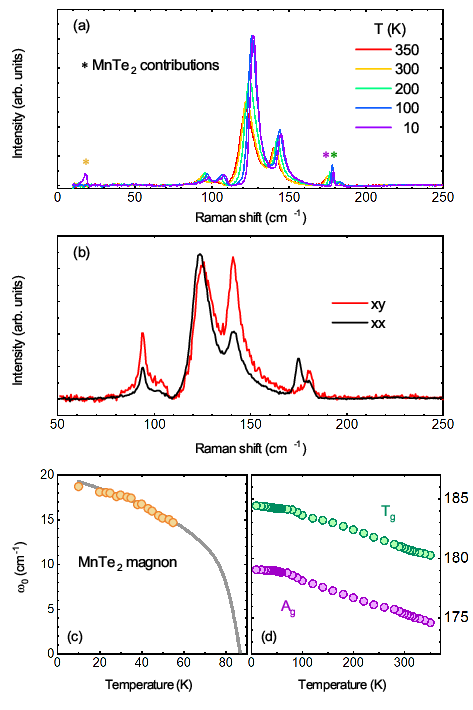}
\caption{\label{fig:R1} (a) Background-subtracted temperature-dependent Raman spectra measured at S1-POS2. Asterisks label the MnTe$_2$-related features that can be identified via their temperature dependence. (b) Polarization-dependent spectra show that the 175\,cm$^{-1}$ mode disappears in cross-polarization, whereas the 180\,cm$^{-1}$ mode weakens in the co-polarization configuration, thus confirming the assignment to the $A_g$ and $T_g$ modes of MnTe$_2$, respectively. 
(c,d) Temperature-dependent resonance frequencies of the MnTe$_2$-related modes. The magnon mode in (c) shows an order parameter-like behavior (solid line) with the critical temperature of $\sim$90~K, which is very close to the N\'eel temperature of 87\,K in MnTe$_2$. Both $A_g$ and $T_g$ modes show weak anomalies around this temperature, while they do not exhibit any abrupt changes at 307~K, the critical temperature for $\alpha$-MnTe. }
\end{figure}


Raman spectra of $\alpha$-MnTe show several modes, in contrast to the single Raman-active $E_g$ phonon expected from the symmetry analysis. It remains unclear which of the experimental modes corresponds to this phonon, and what causes the appearance of the additional modes.
In Fig.~\ref{fig:R}a, we present room-temperature spectra measured on single crystals from two different batches, S1 and S2. Four modes are systematically observed in both crystals. Two intense modes, R3 and R4, appear around 120 and 140\,cm$^{-1}$, respectively, whereas two weaker modes, R1 and R2, can be seen at $90-100$\,cm$^{-1}$. Whereas the same four lines were observed at all positions of S2, measurements at different positions on the surface of S1 revealed two additional modes around 175 and 180\,cm$^{-1}$. A clear visual contrast between the regions where these additional modes appear (POS2) and disappear (POS1) suggests that such modes, often reported in the previous literature~\cite{mobasser1985,zhang2020,wu2025}, originate from surface inclusions of a secondary phase (see the inset of Fig.~\ref{fig:R1}a). Therefore, we also performed the Raman study on these inclusions (S1-POS2) to explore their origin.

\begin{figure}
\includegraphics[width=8.6cm]{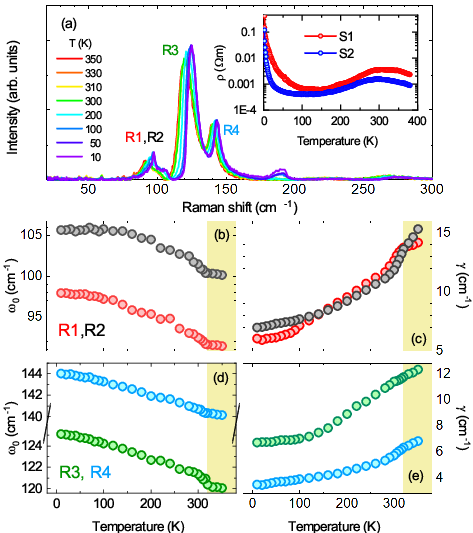}
\caption{\label{fig:R2} (a) Temperature-dependent Raman spectra measured on the sample S2. The inset shows temperature dependence of the electrical resistivity measured on both crystals. (b-e) Resonance frequencies and scattering rates of the modes R1--R4. All modes are affected by the magnetic ordering, and small anomalies are observed across. The additional mode at $\sim$190\,cm$^{-1}$ appearing at low temperatures is tentatively assigned to 2E$_{2g}$. }
\end{figure}

Several observations indicate MnTe$_2$ as the origin of the additional modes at $175-180$\,cm$^{-1}$. First, temperature-dependent Raman spectra measured at S1-POS2 (Fig.\ref{fig:R1}a) display an additional low-energy excitation appearing at $15-20$\,cm$^{-1}$ below 60\,K. Both energy and temperature dependence of this excitation match the known magnon mode of MnTe$_2$~\cite{fahmy2025} with the reported N\'eel temperature of 87\,K~\cite{burlet1997}. No such mode appears at either S1-POS1 or S2. Second, frequencies of the $175-180$\,cm$^{-1}$ modes show a weak anomaly around 90\,K, near the magnetic ordering temperature of MnTe$_2$, but they evolve smoothly at $T_N$ of $\alpha$-MnTe, in contrast to the R1--R4 modes that will be discussed later in this section.


We further investigated the $175-180$\,cm$^{-1}$ modes via polarization-dependent measurements at room temperature. As shown in Fig.~\ref{fig:R1}b, these modes are sensitive to the polarization direction. The 175\,cm$^{-1}$ mode is suppressed in cross-polarization, where the 180\,cm$^{-1}$ mode is enhanced. This behavior is well in line with their anticipated $A_g$ and $T_g$ symmetries (Table~\ref{tab:mnte2}) that allow the $xx$ and $xy$ polarizations, respectively. It is worth noting that Ref.~\cite{wu2025} also reports the disappearance of the 175\,cm$^{-1}$ mode in cross-polarization, but the mode itself is assigned to $\alpha$-MnTe, at odds with our data. 


The final check for the secondary phase as the origin of the $175-180$\,cm$^{-1}$ modes is done by varying the laser power (Fig.~\ref{fig:R}b). With a higher laser power, we observe damage on the crystal surface, accompanied by the appearance of a broad mode around 175\,cm$^{-1}$ along with a few higher-energy modes of unknown origin. This experiment further identifies the $175-180$\,cm$^{-1}$ modes as extrinsic. 

While there is ample evidence for the $175-180$\,cm$^{-1}$ modes not belonging to $\alpha$-MnTe, the four other modes, R1--R4, appear to be intrinsic and well reproducible. We now analyze these modes in more detail using the temperature-dependent spectra measured on the sample S2 (Fig.~\ref{fig:R2}a), which is free from the MnTe$_2$ inclusions. Importantly, all the four modes display anomalies in both frequency and linewidth around $T_N$. This further confirms their intrinsic nature. 

Our DFT calculations place the Raman-active $E_{2g}$ phonon near 100\,cm$^{-1}$ (Table~\ref{tab:mnte}), in agreement with the recent \textit{ab initio} studies~\cite{thapa2026}. This frequency matches the positions of the modes R1 and R2. The two modes are close siblings, as they show an almost identical temperature evolution with the constant offset of 7\,cm$^{-1}$. Such a splitting of the $E_{2g}$ mode could indicate the broken 6-fold symmetry of the structure, although any macroscopic symmetry lowering is clearly excluded by our XRD data (Fig.~\ref{fig:xrd}). One possible explanation for the mode splitting is the local strain caused by laser heating, which is further corroborated by a visible peak shift upon increasing the laser power (Fig.~\ref{fig:R}b). We also note that an additional small peak around 190\,cm$^{-1}$ is clearly visible at low temperatures and gradually broadens upon warming. This peak can be interpreted as the first overtone of $E_{2g}$, although its absence in the sample S1 with a slower drop of the electrical resistivity (Fig.~\ref{fig:R2}a, inset) may also indicate a defect origin. Alternatively, it could be a plasmon excitation discussed in Ref.~\cite{thapa2026}. 


Another interesting feature of the Raman spectra is the appearance of two strong modes at $\sim$120\,cm$^{-1}$ (R3) and $\sim$140\,cm$^{-1}$ (R4), respectively. These modes are observed in every $\alpha$-MnTe sample. Previously, they were attributed to an impurity of elemental tellurium~\cite{mueller1991,szuszkiewicz2014,shao2026} that indeed shows the Raman-active modes $A_1$ and $E$ at 118 and 139\,cm$^{-1}$, respectively~\cite{yuan2022}. However, both the R3 and R4 modes feature anomalies around $T_N$, which would not be the case if elemental tellurium were their origin. Moreover, the 140\,cm$^{-1}$ mode becomes weaker in cross-polarization (Fig.~\ref{fig:R1}b), at odds with the symmetry analysis for elemental tellurium that predicts the suppression of the $A_1$ mode in this configuration. 

\begin{figure}
\includegraphics{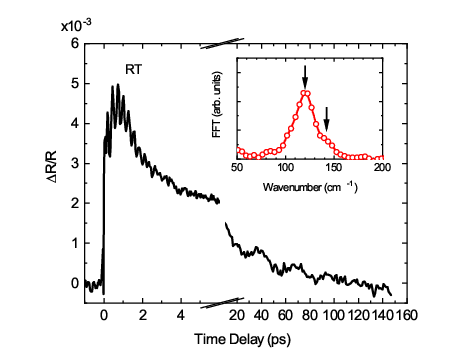}
\caption{\label{fig:PP}Room-temperature transient reflectivity spectrum of $\alpha$-MnTe (sample S2). Short delay times reveal coherent phonon oscillations. The Fourier-transformed signal shown in the inset displays two frequencies that correspond to the R3 and R4 Raman modes, respectively. 
}
\end{figure}

\begin{figure*}
\includegraphics[width=0.95\textwidth]{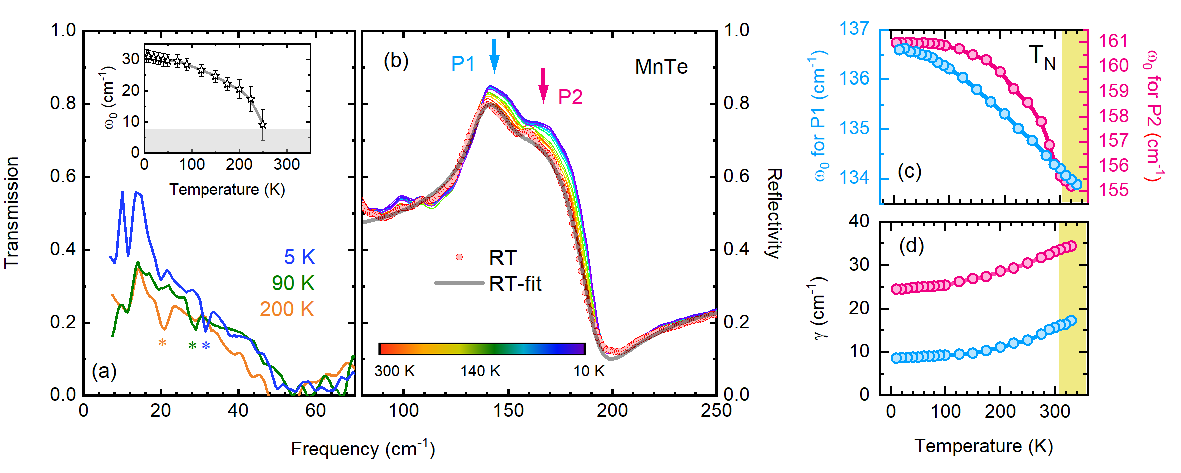}
\caption{\label{fig:IR} (a) Transmission spectra of $\alpha$-MnTe obtained via THz-TDS measurements. Asterisks mark the optical magnon mode, with its temperature dependence presented as an inset. The shaded area in the figure is the experimental low-frequency limit that prevented us from tracing this mode up to room temperature. (b) Temperature-dependent infrared reflectivity of $\alpha$-MnTe. A fit of the room-temperature spectrum is also shown, with two modes, P1 and P2, marked with the arrows. (c,d) Resonance frequencies and scattering rates for the P1 and P2 modes obtained from the reflectivity fit. }
\end{figure*}

One further piece of evidence for the intrinsic nature of the R3 and R4 modes comes from the earlier ultrafast MOKE experiments~\cite{gray2024}. The $\alpha$-MnTe sample was excited with 1560 nm light, and coherent phonon oscillations at 3.6 THz and 4.2 THz (R3 and R4, respectively) were observed.
The coupling of R3 and R4 to the optical pump is intriguing, as it can serve as a tuning mechanism for the optical control of altermagnetic properties. Here, we verified this coupling by transient reflectivity measurements performed with the 800\,nm light for both pump and probe and the fluence of 1.2\,mJ/cm$^2$ [Fig.~\ref{fig:PP}]. The measurements were performed on the sample S2, which should be free from the MnTe$_2$ inclusions. The 3.6 and 4.2\,THz oscillations can be indeed observed with great accuracy. 

It is instructive to compare our transient reflectivity data with the measurements on elemental tellurium~\cite{kamaraju2010} that demonstrated a strong anharmonic damping of the phonon frequency upon increasing fluence. For very low fluences, the main 3.6 THz mode is observed as a coherent oscillation, whereas for fluences on the order of 1\,mJ/cm$^2$, as used in our experiment, a strong damping to 3.3-3.4\,THz occurred. Furthermore, the 4.2 THz mode of elemental tellurium couples to the transient reflectivity in extremely anharmonic cases only. Here, we observed the 3.6 THz mode at the expected energy (no damping), as well as the 4.2\,THz mode using moderate fluence. This comparison further rules out elemental tellurium as the origin of R3 and R4 and indicates their intrinsic nature. We also note that we do not observe a coherent phonon oscillations at 5.3\,THz that would correspond to the 175\,cm$^{-1}$ mode. This is in contrast to Ref.~\cite{bossini2021} that reported such an oscillation in $\alpha$-MnTe using the same 800\,nm pump as in our study. We thus argue that this observation could be caused by the MnTe$_2$ inclusions, similar to the Raman spectra for S1 - POS2, as discussed above.

One plausible explanation for R3 and R4 could be \mbox{non-$\Gamma$} phonon modes that may appear in Raman spectra because of higher-order Raman processes. According to the calculated phonon dispersions~\cite{mu2019}, zone-boundary acoustic modes at $M$, $K$, and $A$ appear at about half of the R3 and R4 energies and may give rise to a two phonon process. Experimental study of the non-$\Gamma$ phonons in \mbox{$\alpha$-MnTe} would be desirable in order to verify this scenario and shed further light on the nature of R3 and R4. 

\subsection{Infrared-active phonons and the low-energy magnon mode }

IR spectroscopy offers complementary information on the phonon modes in $\alpha$-MnTe. 
Previous IR experiments agree on the energy range of the phonon excitations but leave ambiguities in their assignment. From the symmetry analysis, one expects two IR-active phonons, $A_{2u}$ and $E_{1u}$ polarized along the $c$-axis and in the $ab$-plane, respectively. A measurement on a plate-like single crystal is expected to reveal the $E_{1u}$ mode only, because electric field of light lies in the $ab$-plane. On the other hand, both modes could be potentially observed on powder samples. Earlier studies do not specify the exact orientation of the sample and generally report a single phonon mode in the range of $130-154$\,cm$^{-1}$~\cite{povstyanyi1972, onari1974, allen1977, gao2026}. 

For an unambiguous determination of the mode symmetry, we performed temperature-dependent IR spectroscopy measurements on the S2 crystal with $E\parallel ab$-plane (Fig.~\ref{fig:IR}b). A similar measurement was performed on S1 at room temperature. The spot size of the IR measurements is much larger compared to Raman, so the spectrum is representative of the whole crystal, including its POS2 region with MnTe$_2$, but no characteristic phonons of MnTe$_2$ were observed at $150-170$\,cm$^{-1}$. The IR spectra of S1 and S2 are essentially identical. It means that the MnTe$_2$ inclusions are mainly present on the surface. They do not appear in the IR measurements with the large spot size and penetration depth. 

Our room-temperature IR spectra display two modes, P1 and P2, centered around 134 and 155\,cm$^{-1}$, respectively. The P2 mode shows a strong anomaly at the magnetic ordering temperature, with an order-parameter-like behavior below $T_N$, indicating a strong coupling to the spin dynamics. This mode can be assigned to the $E_{1u}$ phonon. Its frequency shows an excellent agreement with the DFT results (SCAN, $U_{\rm eff}=4$\,eV, see Table~\ref{tab:mnte}). Furthermore, the scattering rate of this mode is rather large even at the lowest temperature, suggesting a strong-coupling scenario.

The P1 mode is remarkably different. It is much sharper than P2, and despite a slight anharmonicity, no anomaly at $T_N$ is observed. 
At first glance, and in view of the DFT results, this mode could be assigned to the second IR-active phonon. However, that phonon should have the $A_{2u}$ symmetry with the purely out-of-plane atomic displacements. Given our measurement configuration with the in-plane orientation of the electric field of the IR light, the $A_{2u}$ mode should not be observable in our measurement. 

\begin{figure}
\includegraphics{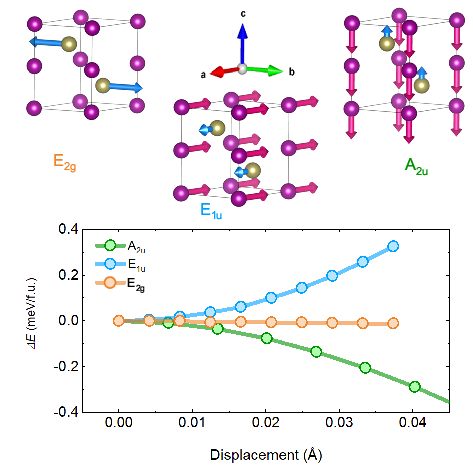}
\caption{\label{fig:SP} Atomic displacements in the IR- and Raman-active phonon modes of $\alpha$-MnTe and the associated spin-phonon coupling gauged by the energy difference between the ferromagnetic and antiferromagnetic states, $\Delta E=E_{\rm FM}-E_{\rm AFM}$, calculated within DFT+$U$ (PBE, $U=5$\,eV, $J=1$\,eV). The energy difference in the equilibrium structure is taken as zero of $\Delta E$. }
\end{figure}

Spin-phonon coupling can be a useful test for the nature of various phonon modes. In Fig.~\ref{fig:SP}, we show how atomic displacements of the IR- and Raman-active modes change the energy difference between the ferromagnetic and antiferromagnetic states of $\alpha$-MnTe. It turns out that the $E_{1u}$ mode stabilizes the antiferromagnetic state, whereas the $A_{2u}$ mode destabilizes it. This is likely because of the opposite changes in the Mn--Te--Mn bond angles that control superexchange interactions in $\alpha$-MnTe. The hardening of the P2 mode below $T_N$ is consistent with the stabilization of magnetic order by the $E_{1u}$ phonon. On the other hand, the $A_{2u}$ mode should soften below $T_N$, whereas P1 shows a very different behavior experimentally (Fig.~\ref{fig:IR}).

Taken together, these observations suggests that the P1 mode should not be assigned to the IR-active $A_{2u}$ phonon, even though its calculated frequency perfectly matches the experimental value. Similar to the R3 and R4 modes, P1 may be caused by a higher-order process involving non-$\Gamma$ phonons. Experimental studies of the full phonon dispersion of $\alpha$-MnTe would be necessary in order to pinpoint its exact origin. It is also worth noting that previous IR and Raman spectroscopy experiments on the sibling compound $\alpha$-MnSe revealed a plethora of combined modes caused by zone-boundary phonons~\cite{popovic2006}.

Finally, we have investigated the THz spectral region using a THz-TDS setup to probe optical magnons. A recent magneto-optical spectroscopy study~\cite{dzian2025} revealed a mode at 3.5 meV at 4 K across different $\alpha$-MnTe samples with varying doping levels. For characterization purposes, we performed a similar study on our S2 sample. Fig.\ref{fig:IR}a shows the anticipated magnon mode and its temperature dependence. Due to experimental limitations, we could not unambiguously detect this mode above 250\,K, but the overall temperature dependence and red shift on warming are consistent with the previous studies~\cite{dzian2025}.  

\section{Discussion and Conclusions}

By combining IR and Raman spectroscopies with high-resolution XRD, we have shown that tiny inclusions of MnTe$_2$ may appear in single crystals of $\alpha$-MnTe and cause some of the experimental signatures that were previously assigned to the altermagnetic phase of the material. The Raman modes at $175-180$\,cm$^{-1}$ are unaffected by the magnetic ordering transition of $\alpha$-MnTe, yet they clearly correlate with the observation of the magnon mode of MnTe$_2$, which renders MnTe$_2$ phonon excitations as their most plausible origin. The alternative explanation in terms of a plasmon mode~\cite{thapa2026} seems less likely because it accounts neither for the spatial dependence nor for the simultaneous appearance/disappearance of the $175-180$\,cm$^{-1}$ modes and the lower-energy mode assigned to the MnTe$_2$ magnon.

We further note that weak ferromagnetism reported below $80-90$\,K in some of the $\alpha$-MnTe samples~\cite{orlova2024,orlova2025} may also originate from the MnTe$_2$ inclusions, because it is not reproducible across different reports either~\cite{roessler2025,wu2025b}. From the experimental perspective, identifying such MnTe$_2$ inclusions is by no means a simple task. A careful mapping of individual crystals using Raman or energy-dispersive x-ray spectroscopy (EDX) seems essential to verify the absence of MnTe$_2$ inclusions.

Another word of caution concerns optical measurements. Bulk probes such as IR spectroscopy are less likely to be affected by secondary phases because of the large spot size and high penetration depth, at least at low energies, which are relevant to phonon and magnon excitations. On the other hand, surface-sensitive probes with the small spot size and low penetration depth can pick up the signals from secondary phases. We foresee that Raman spectroscopy and ultrafast techniques utilizing the pump and/or probe in the visible range are especially vulnerable. 

In conclusion, we have shown that $\alpha$-MnTe retains its $P6_3/mmc$ symmetry within our experimental resolution. Using spatially resolved and temperature-dependent measurements we have revised the mode assignment in the IR and Raman spectra of altermagnetic $\alpha$-MnTe. The $175-180$\,cm$^{-1}$ modes are shown to be extrinsic and caused by the inclusions of MnTe$_2$ on the surface of the crystal. On the other hand, the intense Raman modes at 120 and 140\,cm$^{-1}$, which were previously associated with the impurity of elemental tellurium, are in fact intrinsic. They display clear anomalies at the magnetic ordering transition of $\alpha$-MnTe, as well as coherent phonon oscillations in transient reflectivity that open the way to manipulating $\alpha$-MnTe with the optical pump. The IR and Raman excitations of $\alpha$-MnTe are not restricted to $\Gamma$-point optical phonons and involve more complex mechanisms that require a further dedicated investigation.

\acknowledgments
We thank Ulrich R\"o{ssler} for fruitful discussions on the $\alpha$-MnTe project, Vicky Haase for technical assistance in crystal growth, and Uta Luchessi for her support during the Raman measurements. We acknowledge ESRF for providing the beamtime for this experiment and acknowledge the technical assistance by Andy Fitch during the data collection at ID22. We also acknowledge the computing time made available on the high-performance computer at the NHR Center of TU Dresden. This center is jointly supported by the Federal Ministry of Education and Research and the state governments participating in the NHR [www.nhr-verein.de/unsere-partner]. Work in HZDR is partially supported by DFG (UY63/2 and UY63/5).

\bibliography{MnTe-phonon}

\end{document}